\documentclass{JHEP3}
\usepackage{graphicx}
\usepackage{amsmath,amssymb}
\usepackage{epsfig}
\usepackage{cite}

\title{The extremal black hole bomb}

\author{J. G. Rosa\\ Rudolf Peierls Centre for Theoretical Physics, Department of Physics,\\ University of Oxford, 1 Keble Road, Oxford OX1 3NP, United Kingdom\\ Email: \email{j.rosa1@physics.ox.ac.uk}}

\received{\today}

\abstract{We analyze the spectrum of massive scalar bound states in the background of extremal Kerr black holes, focusing on modes in the superradiant regime, which grow exponentially in time and quickly deplete the black hole's mass and spin. Previous analytical estimates for the growth rate of this instability were limited to the $\mu M\ll1$ and $\mu M\gg1$ regimes, where $\mu$ and $M$ denote the scalar field and black hole masses, respectively. In this work, we discuss an analytical method to compute the superradiant spectrum for generic values of these parameters, namely in the phenomenologically interesting regime $\mu M\sim 1$. To do this, we solve the radial mode equation in two overlapping regions and match the solutions in their common domain of validity. We show that matching the functional forms of these functions involves approximations that are not valid for the whole range of scalar masses, exhibiting unphysical poles that produce a large enhancement of the growth rate. Alternatively, we match the functions at a single point and show that, despite the uncertainty in the choice of the match point, this method eliminates the spurious poles and agrees with previous numerical computations of the spectrum using a continued-fraction method.}

 \preprint{OUTP-09-29-P}

\begin{document}

\section{Introduction}

The issue of black hole stability plays a central role in General Relativity and is crucial for our understanding of the rich high-energy phenomenology associated with compact astrophysical systems. The issue of stability of the Schwarzschild geometry was first addressed in the seminal paper by Regge and Wheeler \cite{Regge:1957td} and later by Vishveshwara \cite{Vishveshwara:1970cc} and Wald \cite{Wald:1979}, showing that small perturbations of this spherically symmetric space were exponentially damped in time. It has been known for a few decades that this is not, however, the case for rotating Kerr black holes \cite{Kerr:1963ud}, which suffer from the well-known superradiant instability \cite{Zeldovich:1971,Zeldovich:1972,Damour:1976kh,Zouros:1979iw,Detweiler:1980uk,Furuhashi:2004jk,Strafuss:2004qc,Cardoso:2005vk,Dolan:2007mj}.

 This instability is a direct consequence of the presence of an {\it ergoregion} surrounding the outer horizon of the black hole, where ordinary causal matter experiences inertial frame-dragging and cannot remain at rest with respect to an asymptotic observer. Within this region, the Killing vector associated with time translations becomes space-like, so that the associated conserved energy may become negative for some observers. Classically, this gives rise to the so-called Penrose process \cite{Penrose:1969pc, Christodoulou:1970wf}, by which one may extract energy and angular momentum from the spinning black hole. One can consider a projectile directed towards the black hole along a time-like trajectory but aimed in such a way that it misses the outer horizon. One could then conceive a timer mechanism that sets off a fragmentation process that breaks the projectile into two parts while it is inside the ergoregion. One of the fragments may then carry a negative energy and fall into the black hole, while the remaining fragment escapes to infinity carrying an energy larger than that of the original fragment. One can also show that the same reasoning holds for the angular momentum of both fragments.
 
Superradiance is a well-known phenomenon in several quantum \cite{Manogue, Greiner} and also classical systems \cite{Zeldovich:1971mw} and was first described for the case of spinning black holes by Zeldovich \cite{Zeldovich:1971, Zeldovich:1972}. If one considers the scattering of wave modes of the form $e^{-i\omega t+im\phi}$, where $\phi$ denotes the Boyer-Lindquist azimuthal angle in the Kerr geometry \cite{Boyer:1966qh}, one can show that for co-rotating ($m>0$) incident waves satisfying the condition $\omega<m\Omega$, where $\Omega={a\over r_+^2+a^2}$ is the angular velocity of the black hole at the outer horizon $r_+$, the scattered wave is amplified. The total energy and angular momentum radiated to infinity are hence larger than that of the incident wave, in a similar fashion to what happens in the Penrose process described above. Press and Teukolsky proposed that, by surrounding the black hole with a mirror, one could use multiple wave-scattering to extract an enormous amount of energy from the black hole, a mechanism they suggestively named the {\it black hole bomb} \cite{Teukolsky} and that was further analyzed in \cite{Cardoso:2004nk}.
 
Practical implementations of such a mechanism seem difficult to conceive, although it has been proposed in \cite{Putten, Aguirre:1999zn} that the inner boundary of the accretion disks which are expected to surround astrophysical black holes may effectively act as a mirror and produce enough energy via superradiant scattering of magnetosonic plasma waves to feed the mysterious and highly-energetic gamma-ray bursts. A natural mirror is, however, provided for fields with a non-zero mass \cite{Damour:1976kh, Zouros:1979iw}. It is well-known that stable orbits around a black hole are an exclusive feature of particles with a non-vanishing mass, so that massive fields exhibit a set of bound states in the Kerr background corresponding to wave-packets moving along such orbits. These wave-packets will always have a non-vanishing tail inside the black hole's ergoregion, leading to a continuous amplification of bound states satisfying the superradiant condition. Such states have a complex frequency with a small positive imaginary part, so that their occupation number grows exponentially in time. Pauli blocking prevents this amplification for fermionic states \cite{Unruh}, so that superradiant bound states are an exclusive feature of bosonic fields.

Several attempts have been made in the literature to analytically determine the growth rate of superradiant bound states for massive scalar fields. The difficulty in solving the massive Klein-Gordon equation in the Kerr background has limited the analysis to the regimes $M\mu\gg1$  \cite{Zouros:1979iw} and $M\mu\ll1$ \cite{Detweiler:1980uk, Furuhashi:2004jk}, where $M$ and $\mu$ denote the masses of the black hole and the field, respectively, in geometrized units such that $G=c=\hbar=1$. In both cases the obtained growth rates were shown to be quite small, with the imaginary part of the frequency $\omega_I M$ scaling as $e^{-1.84 \mu M}$ and $(\mu M)^9$ in the large and small mass limits, respectively. These studies suggest as well that the growth rate is maximal for extremal holes with $a=M$ and for P-waves with $l=m=1$. The spectrum of massive scalar bound states has also been studied numerically in \cite{Cardoso:2005vk} and later in \cite{Dolan:2007mj}, using a continued-fraction method which confirmed these expectations. The latter analysis revealed a maximum growth rate in the $l=m=1$ case given by $\omega_I M\simeq 1.5\times10^{-7}$, for $\mu M\simeq0.42$ and $a=0.999M$, in agreement with the former, where a similar value was observed for $a=0.994$ and $\mu M=0.45$.

Hence, numerical results suggest that the superradiant instability is more pronounced in the regime $\mu M\sim 1$, which has proven hard to analyze with analytical methods, motivating a recent analytical study of this phenomenon \cite{Hod:2009cp}. By separating the exterior of the black hole into two overlapping regions where the field equations are exactly solvable and matching the obtained solutions within their common domain of validity, the authors obtained a defining equation for the superradiant bound state spectrum valid for $\mu M\sim 1$. The novelty of their method resides in the use of the black hole's angular momentum rather than the particle's mass to define the two regions, thus eliminating the need for large and small mass approximations in order to solve the relevant equations. The results in \cite{Hod:2009cp} exhibit, however, a huge discrepancy with respect to the numerical analysis in \cite{Cardoso:2005vk, Dolan:2007mj}, with a maximum growth rate of $\omega_I M\simeq 1.7\times10^{-3}$, four orders of magnitude larger than the one mentioned above.

Understanding this discrepancy is crucial for the study of the phenomenological aspects of scalar superradiant bound states in the Kerr background, in particular as this process may lead to the significant spin down of rapidly-rotating black holes. Astrophysical black hole candidates have been primarily found in X-ray binaries and  Active Galactic Nuclei (AGN), with masses in the range $3M_{\odot}-30M_{\odot}$ for stellar mass black holes and $10^6M_{\odot}-10^9M_{\odot}$ for supermassive black holes, where $M_{\odot}$ denotes the solar mass \cite{McClintock:2009as}. The lightest known scalar particle is the neutral pion, with $\mu=134.98$ MeV and a mean lifetime $\tau_\pi\simeq 8.4\times10^{-17}$ sec \cite{Amsler:2008zzb}, so that it may only significantly affect small black holes with a mass of about $10^{12}$ kg, given the requirement $\mu M\sim 1$. Hence, pionic bound states are not expected to produce observable effects on astrophysical systems, although they may influence the dynamics of small primordial black holes postulated to be produced during inflation \cite{Zouros:1979iw}. This statement depends crucially on whether the instability can develop significantly before the pions decay. While a large amplification of pionic bound states can be attained using the analytical results of \cite{Hod:2009cp}, where the time scale for the instability is $\tau\sim1.3\times10^{-21}$ sec, the numerical results of \cite{Cardoso:2005vk, Dolan:2007mj} yield $\tau\sim1.5\times10^{-17}$ sec, in which case superradiant pion emission would be very ineffective.

Recently, the authors of \cite{Arvanitaki:2009fg} considered the many pseudoscalar moduli fields with axion-like properties which generically appear in string theory compactifications. These moduli arise as Kaluza-Klein zero-modes of the Neveu-Schwarz and Ramond-Ramond closed string antisymmetric form fields and, if massless at tree level,  may acquire masses from non-perturbative effects such as string instantons. The authors then argued that, if realistic compactifications are to include the QCD axion as a solution to the strong CP problem, string instanton contributions to its potential must be suppressed with respect to non-perturbative QCD contributions by at least a factor of $10^{10}$. This suggests that many of the axion-like moduli will be lighter than the QCD axion, generating a plethora of ultra-light scalar states they denoted as the ``string axiverse", with masses possibly down to the Hubble scale, $H_0\sim10^{-33}$ eV. Many of these axions may thus lie in the correct mass range to significantly spin down astrophysical black holes, possibly leading to several interesting observational effects such as axion-photon conversion, through the coupling to the surrounding magnetic fields, and emission of gravitational waves, via quantum mechanical transitions between superradiant and non-superradiant states. 

Spinning astrophysical black holes may thus provide a simple probe of string theory compactifications. In particular, black holes with mass $M\sim \mu^{-1}$ for a given axion should exhibit a very low angular momentum, so that looking for gaps in the mass-spin Regge plot for astrophysical black holes may provide an effective method to search for ultra-light states  \cite{Arvanitaki:2009fg}. However, the shape of these gaps will be extremely sensitive to the variation of the superradiant growth rate with the scalar mass. Moreover, black holes may also gain angular momentum from the surrounding accretion disks, a process that may compete with the superradiant spin down if the instability does not develop sufficiently fast.

The rich phenomenology associated with the superradiant instability thus requires a more detailed study of its properties and an analysis of the above mentioned discrepancy. With this goal in mind, in this paper we compute the spectrum of superradiant massive scalar bound states in an extremal Kerr background, for which the growth of the instability is expected to be maximal. We consider a functional matching procedure analogous to the one used in \cite{Hod:2009cp} and show that, unfortunately, it is not applicable for the whole scalar mass range, mainly due to the approximations involved. Alternatively, we consider a point matching procedure and show that, within the uncertainty associated with the choice of the match point, this method agrees with the numerical results of \cite{Dolan:2007mj} for a large range of masses, in particular in the vicinity of the above mentioned discrepancy. 

This work is organized as follows. In the next section, we analyze the main properties of massive Klein-Gordon fields in a maximally-rotating black hole. Sections III and IV are devoted, respectively, to the analyses of the functional and point matching procedures. We conclude, in Section V, with a discussion of the main advantages and disadvantages of both techniques and some prospects for future work.


\section{The Klein-Gordon equation in an extremal Kerr background}

We start by considering the Klein-Gordon equation in the Kerr background, which for a scalar field $\Phi$ of mass $\mu$ can be written as:
\begin{eqnarray} \label{Klein-Gordon}
(\nabla^{\mu}\nabla_{\mu}-\mu^2)\Phi=0~.
\end{eqnarray}
In terms of the Boyer-Lindquist coordinates $(t,r,\theta,\phi)$  \cite{Boyer:1966qh}, the scalar field admits a mode expansion of the form:
\begin{eqnarray} \label{mode_expansion}
\Phi_{lm}(t,r,\theta,\phi)=e^{-i\omega t}R_{lm}(r)S_{lm}(\theta)e^{i m\phi}~,
\end{eqnarray}
 where $\omega=\omega_R+i\omega_I$ is the complex frequency of the mode characterized by the indices $l = 0,1,\ldots$ and $-l\leq m\leq l$. The latter correspond to the usual angular momentum and azimuthal projection indices for spherical harmonics in the Schwarzschild limit $a\rightarrow 0$. In the general case, they label the solutions of the spheroidal harmonic equation \cite{Abramowitz} corresponding to the angular part of Eq. (\ref{Klein-Gordon}):
\begin{eqnarray} \label{angular_eq}
\bigg[\frac{1}{\sin\theta}\frac{d}{d\theta}\bigg(\sin\theta\frac{d}{d\theta}\bigg)-a^2q^2\cos^2\theta-\frac{m^2}{\sin^2\theta}\bigg]S_{lm}=A_{lm}S_{lm}~,
\end{eqnarray}
where  $q=\sqrt{\mu^2-\omega^2}$ and $A_{lm}$ is the angular eigenvalue, corresponding to the separation constant relating the radial and angular parts of the Klein-Gordon equation. The regularity of the solution of Eq. (\ref{angular_eq}) at the poles $\theta=0,\pi$ determines a discrete set of angular eigenvalues, which for $a^2q^2\ll1$ are given by an expansion of the form:
\begin{eqnarray} \label{angular_eigenvalues}
A_{lm}=l(l+1)+\sum_{k=1}^{\infty}c_{klm}a^{2k}q^{2k}~.
\end{eqnarray}
We are interested in studying bound states for which $\omega$ is only slightly smaller than $\mu$, i.e. $q\ll1$, so that it will be sufficient to consider the first few terms in this expansion, which can be found in \cite{Abramowitz}. The radial Teukolsky equation is given by \cite{Press:1973zz}:
\begin{eqnarray} \label{radial_eq}
{d\over dr}\bigg(\Delta{dR_{lm}\over dr}\bigg)+\bigg[{K^2\over\Delta}-a^2\omega^2+2ma\omega-\mu^2r^2-A_{lm}\bigg]R_{lm}=0~,
\end{eqnarray}
where $\Delta=r^2+a^2-2Mr$ and $K=(r^2+a^2)\omega-am$. The zeros of $\Delta$ at $r_{\pm}=M\pm\sqrt{M^2-a^2}$ correspond to the outer and inner horizons of the Kerr black hole.

For extremal Kerr black holes, with $a=r_+=r_-=M$, it is useful to introduce the variable:
\begin{eqnarray} \label{x_variable}
x\equiv {r-r_+\over r_+}={r\over M}-1~,
\end{eqnarray}
and the re-scaled function $\Psi_{lm}(x)=xR_{lm}(x)$, which satisfies the Schr\"odinger-like equation:
\begin{eqnarray} \label{psi_eq}
{d^2\Psi_{lm}\over dx^2}+\big[\omega^2-V(\omega,x)\big]\Psi_{lm}=0~,
\end{eqnarray}
where 
\begin{eqnarray} \label{V_potential}
V(\omega,x)=-{4\varpi^2\over x^4}-{8\omega\varpi\over x^3}+{\beta(\beta-1)\over x^2}-{2q\nu\over x}+\mu^2
\end{eqnarray}
and we have defined:
\begin{eqnarray} \label{quantities1}
\nu\equiv {2\omega^2-\mu^2\over q}~,\qquad \varpi\equiv \omega-m\Omega~,
\end{eqnarray}
\begin{eqnarray}
\beta(\beta-1)\equiv A-7\omega^2+\mu^2~,
\end{eqnarray}
where all quantities are measured in units of $M$, $\Omega=1/2$ in the extremal case and we have omitted the indices $l,m$ for simplicity. Without loss of generality, we use the following solution for the coefficient $\beta$:
\begin{eqnarray} \label{beta}
 \beta={1\over2}\bigg(1+\sqrt{1+4A-28\omega^2+4\mu^2}\bigg)\simeq l+1~,
\end{eqnarray}
where we have taken the limit $\omega\sim\mu\ll1$ in the last expression, explicitly showing that $\beta$ labels the angular momentum of the solution. It is a particular feature of the extremal case that one can readily write the radial equation in this form without using the tortoise coordinate, defined by \cite{Zouros:1979iw}:
\begin{eqnarray} \label{tortoise}
{dy\over dx}={r^2+a^2\over \Delta}\qquad\Rightarrow\qquad
 y=x+2\log x-{2\over x}~.
\end{eqnarray}

Despite its simple form, the potential determining the dynamics of the radial function depends on the mode frequency, making this problem hard to solve exactly. To better understand the form of this potential, we use the solution for the real part of the mode frequency that we will obtain later on in Eq. (\ref{H_spectrum}). This is illustrated in Figure 1 for $l=m=1$ and $\mu M=0.4$.
\FIGURE{\epsfig{file=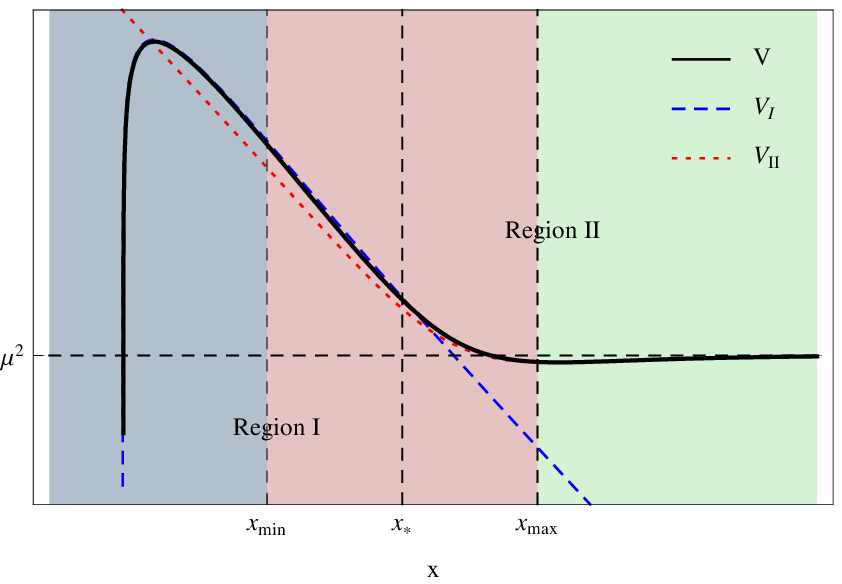, width=9cm}
		\caption{Potential for $l=m=1$ and $\mu M=0.4$ (solid line). Region I corresponds to $x<x_{max}=(2q)^{-1}$, while region II is defined by $x>x_{min}=-4\varpi$. The point $x_{*}=\sqrt{-2\varpi/q}$ corresponds to the trial match point discussed in Section IV. The dashed and dotted curves give approximate forms of the potential valid in regions I and II, respectively.}
		\label{Figure 1}}

The potential can be divided into (i) a sharp potential close to the horizon, (ii) a finite centrifugal barrier and (iii) a potential well. In terms of the tortoise coordinate, the sharp part of the potential tends to a constant value $2m\Omega\omega-m^2\Omega^2$ close to the horizon, which is negative for superradiant modes \cite{Arvanitaki:2009fg}. The definitions of regions I and II, as well as the particular values of $x$ shown in Figure 1, will become clear in the next sections, but they roughly separate the interior and the exterior of the ergoregion. 

The physical picture behind the superradiant amplification of the modes has a simple interpretation in terms of this potential. The Hydrogen-like bound states of the potential well may ``anti-tunnel" \cite{Arvanitaki:2009fg} into the ergoregion and, while a fraction may penetrate the black hole's horizon, the remaining part will be reflected with a larger amplitude for $\omega<m\Omega$. This energy remains localized in the potential well rather than being radiated to infinity, as the wave packets are bound to the black hole, effectively inducing multiple scatterings as if a mirror was placed around the horizon. This results in a small positive imaginary part of the frequency, making the occupation number of the bound states grow exponentially according to Eq. (\ref{mode_expansion}). This effect should be smaller for higher multipoles, for which the barrier height increases and the ``tunneling amplitude" is suppressed.

In order to determine the spectrum of bound states, one needs to impose boundary conditions. These correspond to an ``ingoing wave" at the horizon and an exponentially decaying mode at infinity \cite{Zouros:1979iw, Furuhashi:2004jk}, which in terms of the tortoise coordinate defined in Eq. (\ref{tortoise}) can be written as:
\begin{eqnarray} \label{boundary_conditions}
\Psi_{lm}(y)\rightarrow\begin{cases}
{1\over y}e^{-i\varpi y}, &y\rightarrow -\infty \\
e^{-qy}, &  y\rightarrow +\infty
\end{cases}~.
\end{eqnarray}
Recall that $y\sim -{2\over x}$ and $y\sim x$ close to the horizon and at infinity, respectively. One can rewrite the radial equation in a more convenient form by considering the function $F_{lm}(x)=x^{-\beta} \Psi_{lm}(x)$, which satisfies
\begin{eqnarray} \label{F_eq}
\bigg[x^2{d^2\over dx^2}+2\beta x{d\over dx}+{4\varpi^2\over x^2}+{8\omega\varpi\over x}+2q\nu x-q^2x^2\bigg]F=0~.
\end{eqnarray}
 This cannot be solved exactly but admits approximate solutions in two different regions. These may be matched in a common domain of validity to yield the spectrum of bound states with the boundary conditions in Eq. (\ref{boundary_conditions}), which will be the topic of the remainder of this paper.


\section{Functional matching}

As mentioned in the previous section, Eq. (\ref{F_eq}) can be simplified in two distinct regions by noting that the first two terms in the potential may be neglected for $-4\varpi/ x\ll 1$, where $\varpi<0$ for  superradiant modes, while the last two terms may be neglected for $2qx\ll 1$. This defines the regions I and II depicted in Figure 1 such that $x<x_{max}=(2q)^{-1}$ and $x>x_{min}=-4\varpi$, respectively. In Figure 1 we have also plotted the approximate forms of the potential obtained by neglecting the relevant terms in each region, denoted by $V_I(x)$ and $V_{II}(x)$.

If one is able to determine solutions of the radial equation in these regions, one may determine the spectrum by matching the obtained functions in the overlap region, $x_{min}\ll x\ll x_{max}$. The existence of such a region requires $-8\varpi q<1$, a condition that we cannot {\it a priori} ensure to be satisfied without computing the spectrum. We will nevertheless assume that this is the case and use it as a consistency check of the matching methods.

The equations in both regions can be reduced to the confluent hypergeometric equation, although in region I this requires a change of variable to $y=-2/x$. Note that this variable corresponds to the near-horizon form of the tortoise coordinate, so that we use the same notation for both to emphasize this fact. Also, as $y\simeq x$ for large $x$, we can say that in both regions the radial equation reduces to the confluent hypergeometric equation in the corresponding limit of the tortoise coordinate. This reduction requires writing $F(y)=e^{\alpha y}G(x)$ with an appropriate choice of the complex parameter $\alpha$  in each case. After imposing the boundary conditions Eq. (\ref{boundary_conditions}), we obtain the following solutions:
\begin{eqnarray} \label{solutions}
F_I(x)&=&A_Ie^{2i\varpi\over x}U\bigg(1-\beta-2i\omega,2-2\beta,-{4i\varpi\over x}\bigg)~,\nonumber\\
F_{II}(x)&=&A_{II}e^{-q x}U(\beta-\nu,2\beta,2qx)~,
\end{eqnarray}
where $A_{I,II}$ are normalization constants and $U(a,b,z)$ is the confluent hypergeometric function which is regular as $|z|\rightarrow\infty$. In the overlap region, we may take the $|z|\rightarrow 0$ limit of the latter, given by \cite{Abramowitz}:
\begin{eqnarray} \label{U_limit1}
U(a,b,z)\sim {\pi\over\sin(\pi b)}\bigg[{1\over\Gamma[1+a-b]\Gamma[b]}-{z^{1-b}\over\Gamma[a]\Gamma[2-b]}\bigg]~.
\end{eqnarray}
This yields for the corresponding $\Psi$ functions:
\begin{eqnarray} \label{psi_overlap}
\Psi_I(x)&=&-A_I{\pi\over\sin(2\pi\beta)}\bigg[{x^\beta\over\Gamma[\beta-2i\omega]\Gamma[2-2\beta]}
-{(-4i\varpi)^{2\beta-1}x^{1-\beta}\over\Gamma[1-\beta-2i\omega]\Gamma[2\beta]}\bigg]~,\nonumber\\
\Psi_{II}(x)&=&A_{II}{\pi\over\sin(2\pi\beta)}\bigg[{x^\beta\over\Gamma[1-\beta-\nu]\Gamma[2\beta]}
-{(2q)^{1-2\beta}x^{1-\beta}\over\Gamma[\beta-\nu]\Gamma[2-2\beta]}\bigg]~.
\end{eqnarray}
Hence, both functions have a common functional form in the overlap region, and one may match the coefficients of the $x^{\beta}$ and $x^{1-\beta}$ terms to obtain the following condition:
\begin{eqnarray} \label{matching1}
{\Gamma[1-\beta-\nu]\over\Gamma[\beta-\nu]}=
(-8i\varpi q)^{2\beta-1}{\Gamma[\beta-2i\omega]\over\Gamma[1-\beta-2i\omega]}\bigg({\Gamma[2-2\beta]\over\Gamma[2\beta]}\bigg)^2~.
\end{eqnarray}
For the matching to be possible, $-8\varpi q<1$ and, as in the small mass limit $2\beta-1\simeq 2l+1>0$, we conclude that the RHS of this condition is suppressed, so that the LHS should vanish to leading order, which implies:
\begin{eqnarray} \label{H_spectrum}
\beta-\nu=-n~\Rightarrow\qquad \omega\approx \mu\bigg(1-{(\mu M)^2\over 2(l+1+n)^2}\bigg)~,
\end{eqnarray}
where $n$ is a non-negative integer. This reproduces the Hydrogen-like spectrum obtained in previous works for $\mu M\ll 1$ \cite{Furuhashi:2004jk, Dolan:2007mj}, although significant deviations are observed for $\mu M\sim1$, where corrections to the real part of the frequency arise both from the polynomial equation $\beta-\nu=-n$ and from the finiteness of the RHS of Eq. (\ref{matching1}). This expression confirms nevertheless our physical intuition about the Hydrogen-like nature of the massive bound states in the black hole's potential well.

We need to go beyond this leading approximation to obtain the imaginary part of the spectrum and with this purpose we have used {\it Mathematica} to numerically obtain the roots of Eq. (\ref{matching1}). The {\it FindRoot} command can be used for this purpose, requiring an initial value for which we considered the relevant root of $\beta-\nu=-n$ in each case. In Figure 2, we plot the results for $l=m=1$ and $n=0$, for which we expect the maximum growth rate.

In Figure 2(a), one observes a smooth increase of the growth rate with $\mu$, with a sharp decrease close to the endpoint of the spectrum, where the superradiant condition is no longer satisfied and which for $\mu M\ll1$ lies close to $m/2$. Although this is in agreement with the numerical results of \cite{Cardoso:2005vk, Dolan:2007mj}, one also observes two peak-like structures separated by a sharp decrease in $\omega_I M$. One of these ``peaks" was also observed in \cite{Hod:2009cp} and is behind the four orders of magnitude discrepancy between the two types of analysis. Figure 2(b) depicts a detail of this feature and should be compared with Figure 1 in \cite{Hod:2009cp}.
\FIGURE{ \epsfig{file=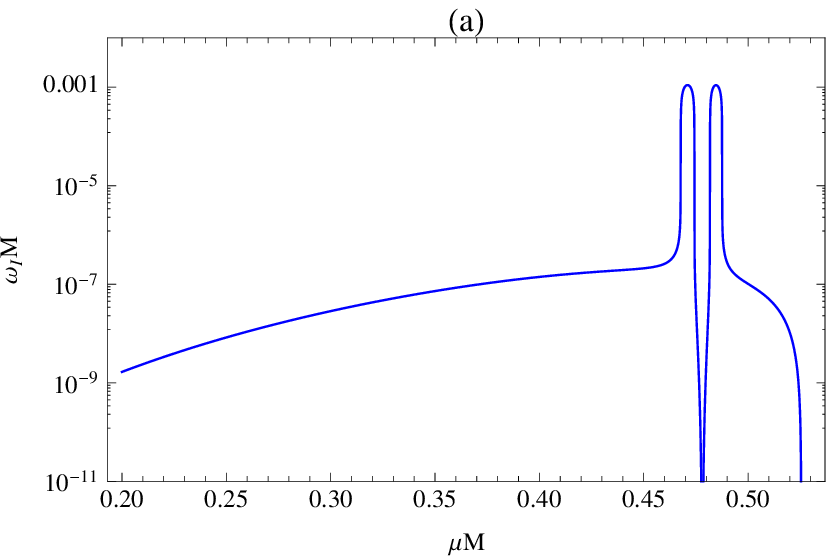,width=8cm}
		\epsfig{file=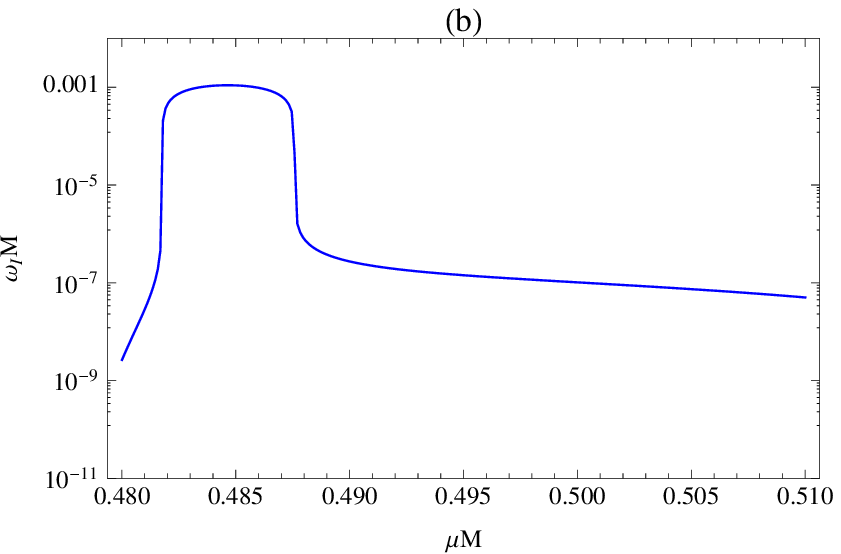,width=8cm}
		\caption{Results obtained for the imaginary part of the frequency obtained using the functional matching approach for $l=m=1$ and $n=0$, illustrating (a) the smooth part and the two peaks and (b) a detail of the second peak.}
		\label{Figure 2}}
One should note, however, that the matching condition Eq. (\ref{matching1}) involves several gamma functions, some of which may develop poles for some values of $\mu$. In particular, $\Gamma[2-2\beta]$ has poles for $\beta=1+p/2$, where $p$ is a non-negative integer. Using the lowest order result in Eq. (\ref{H_spectrum}) for $\omega_R$, one obtains poles in this function for:
\begin{eqnarray} \label{poles}
\mu^2={(2l+1)^2-(p+1)^2\over 24}~,
\end{eqnarray}
which gives $\mu\simeq 0.46$ for $l=p=1$, in the region where the peak-like features in the superradiant spectrum are observed. The shape of this pole is illustrated in Figure 3, showing that it occurs exactly at the value of $\mu M$ where the sharp decrease in the growth rate is observed. This suggests that the peak-like structures are most likely unphysical and simply a consequence of the approximations involved in deriving Eq. (\ref{matching1}).
\FIGURE{	\epsfig{file=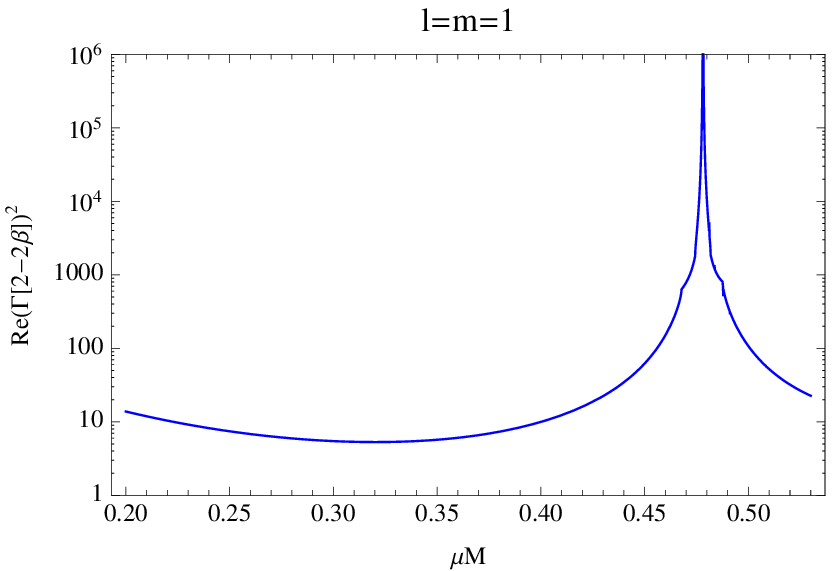, width=9cm}
		\caption{Real part of the function $(\Gamma[2-2\beta])^2$ appearing in the matching condition Eq. (\ref{matching1}), illustrating the pole at the value of $\mu M$ where the sharp decrease in the growth rate of the instability is obtained using the functional matching method.}
		\label{Figure 3}}

In fact, the limit taken in Eq. (\ref{U_limit1}) is not valid close to these poles. To see this, we may rewrite the solution in region I as:
\begin{eqnarray} \label{solution_region_I}
F_I(x)=A_Ie^{2i\varpi\over x}\bigg(-{4i\varpi\over x}\bigg)^{2\beta-1}U\bigg(\beta-2i\omega,2\beta,-{4i\varpi\over x}\bigg)~,
\end{eqnarray}
where we used that  \cite{Abramowitz}:
\begin{eqnarray} \label{aux}
U(a,b,z)=z^{1-b}U(1+a-b,2-b,z)~.
\end{eqnarray}

In this form, we have $b=p+2$ for both solutions, $p\in\mathbb{Z}^+_0$, so that the correct $|z|\rightarrow0$ limit of the confluent hypergeometric function is given by \cite{Abramowitz}:
\begin{eqnarray} \label{U_limit2}
U(a,b,z)\sim {(-1)^b\big[\log z +\psi(a) +\gamma -\psi(b)\big]\over\Gamma[b]\Gamma[a-b+1]}
+{\Gamma[b-1]\over\Gamma[a]}z^{1-b}~,
\end{eqnarray}
where $\psi(x)$ denotes the Digamma function and $\gamma$ the Euler-Mascheroni constant. Using this expression, the solutions in the two regions become:
\begin{eqnarray} \label{psi_overlap_log}
\psi_I(x)&=&A_I\bigg[-\frac{(4i\varpi)^{2\beta-1}\big[\log(-4i\varpi/ x)+\psi(\beta-2i\omega)+\gamma-\psi(2\beta)\big]}{\Gamma[2\beta]\Gamma[1-\beta-2i\omega]}x^{1-\beta}+\frac{\Gamma[2\beta-1]}{\Gamma[\beta-2i\omega]}x^{\beta}\bigg]\nonumber\\
\psi_{II}(x)&=&A_{II}\bigg[{(-1)^{2\beta}\big[\log(2qx)+\psi(\beta-\nu)+\gamma-\psi(2\beta)\big]\over\Gamma[2\beta]\Gamma[1-\beta-\nu]}x^{\beta}+{(2q)^{1-2\beta}\Gamma[2\beta-1]\over\Gamma[\beta-\nu]}x^{1-\beta}\bigg].\nonumber\\
\end{eqnarray}
Hence, the functions cannot be matched in this case, as the $\log x$ corrections are associated with distinct powers of $x$ in each solution. Such logarithmic terms are generically significant in the vicinity of points for which $\beta=1+p/2$, so that the functional matching procedure outlined in this section should only be used away from the above mentioned poles. 

In particular, one cannot {\it a priori} neglect the effects of the logarithmic corrections near the peaks illustrated in Figure 2, even though all gamma functions are finite in this case, due to their obvious proximity to the pole. Furthermore, the discrepancy with respect to the numerical results of \cite{Cardoso:2005vk, Dolan:2007mj} suggests that these corrections are indeed significant. The number of such poles increases as we consider higher multipoles, making it hard to trust this technique in general. This is a disappointing result, as it precludes the use of the functional matching method to cover the whole range of scalar masses.


\section{Point matching}

As we have seen in the previous section, the log-corrections preclude the matching of the functions in the overlap between regions I and II. This is not, however, the only possible technique one can use to obtain the spectrum, and a simple alternative consists in matching the functions and their first derivatives at a point. This technique is widely used in problems in which the potential has two clearly separated regions where the solutions take distinct forms. In this case, however, we have two overlapping regions where one can find approximate solutions to the radial equation. Nevertheless, choosing a match point in the overlap of regions I and II, we may ensure that both functions are sufficiently close to the exact solution and thus obtain a good approximation to the spectrum. One cannot, however, hope to obtain an accurate prediction for the imaginary part of the frequency, as this would require an exact prescription for the choice of the match point, which we do not have {\it a priori}.

Matching the functions in Eq. (\ref{solutions}) at a generic point $x$, one obtains the following condition:
\begin{eqnarray} \label{match_point_generic}
1+2(1-\beta-2i\omega)\frac{U(2-\beta-2i\omega,3-2\beta,-4i\varpi/x)}{U(1-\beta-2i\omega,2-2\beta,-4i\varpi/x)}=\nonumber\\
=\frac{qx^2}{2i\varpi}\bigg[1+2(\beta-\nu)\frac{U(\beta-\nu+1,2\beta+1,2qx)}{U(\beta-\nu,2\beta,2qx)}\bigg]~,\nonumber\\
\end{eqnarray}
where we have used that $U'(a,b,z)=-aU(a+1,b+1,z)$ \cite{Abramowitz}. A good trial match point is $x_*=\sqrt{-2\varpi/q}$, as this corresponds to the value of $x$ for which the absolute values of the arguments of the confluent hypergeometric functions in both regions are the same, simultaneously giving the geometric mean of the values of $x_{min}$ and $x_{max}$ defined in Section II, which give the boundaries of the two regions. This ensures that the matching condition $x_{min}<x<x_{max}$ is satisfied if such an overlap region can be found. We then obtain for this particular choice:
\begin{eqnarray} \label{match_point}
1+2(1-\beta-2i\omega){U(2-\beta-2i\omega,3-2\beta,iz_*)\over U(1-\beta-2i\omega,2-2\beta,iz_*)}=i\bigg(1+2(\beta-\nu){U(\beta-\nu+1,2\beta+1,z_*)\over U(\beta-\nu,2\beta,z_*)}\bigg)\nonumber\\
\end{eqnarray}
with $z_*=\sqrt{-8\varpi q}<1$ for an overlap region to exist. We have determined the roots of this condition numerically for $l=m=1,~2$ and $3$, which give the largest growth rates, via the same procedure used for the functional matching method in the previous section. These results are plotted in Figure 4, along with the corresponding solutions obtained using functional matching and the numerical continued-fraction method in \cite{Dolan:2007mj} for the case $a=0.999 M$, which corresponds to the largest black hole spin analyzed in the latter work. As mentioned above, there is no absolute prescription for the choice of the match point, so that we also plot in this figure the results obtained by choosing different points within the overlap region, $x_{min}<x<x_{max}$, giving an effective uncertainty for the match point method.

\FIGURE{
	\epsfig{file=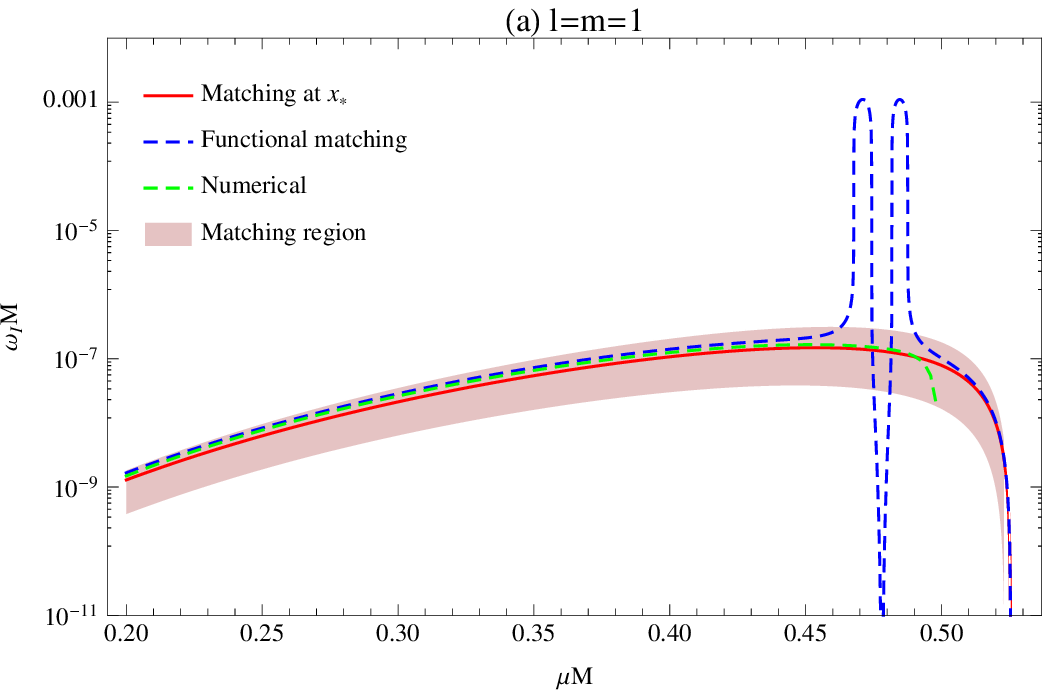, width=9.31cm}
	\epsfig{file=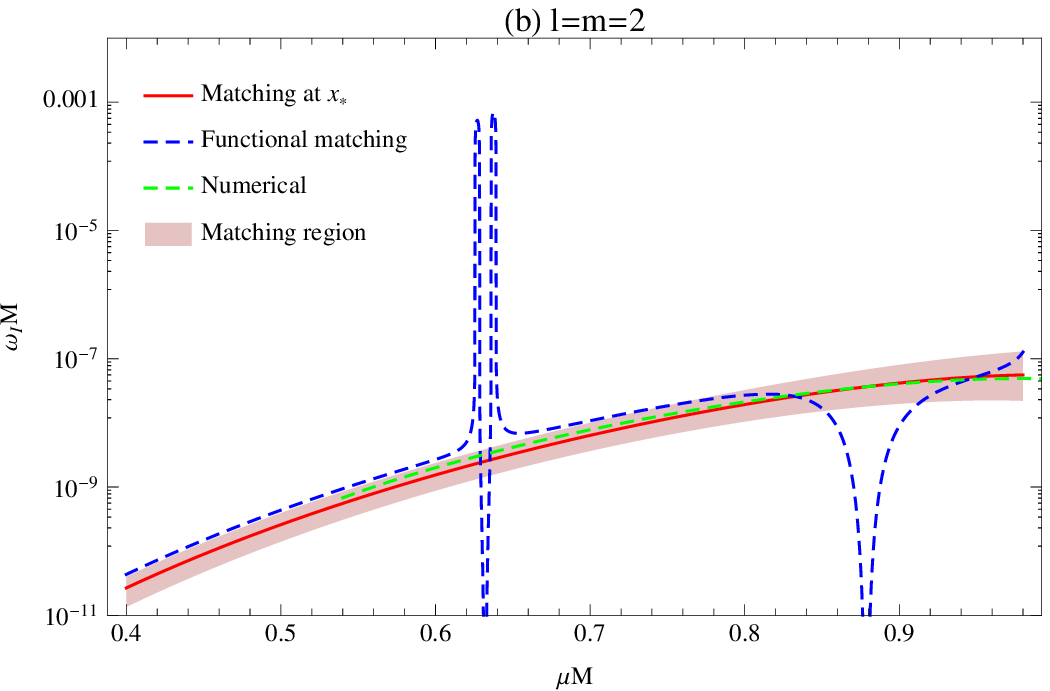, width=9.31cm}
	\epsfig{file=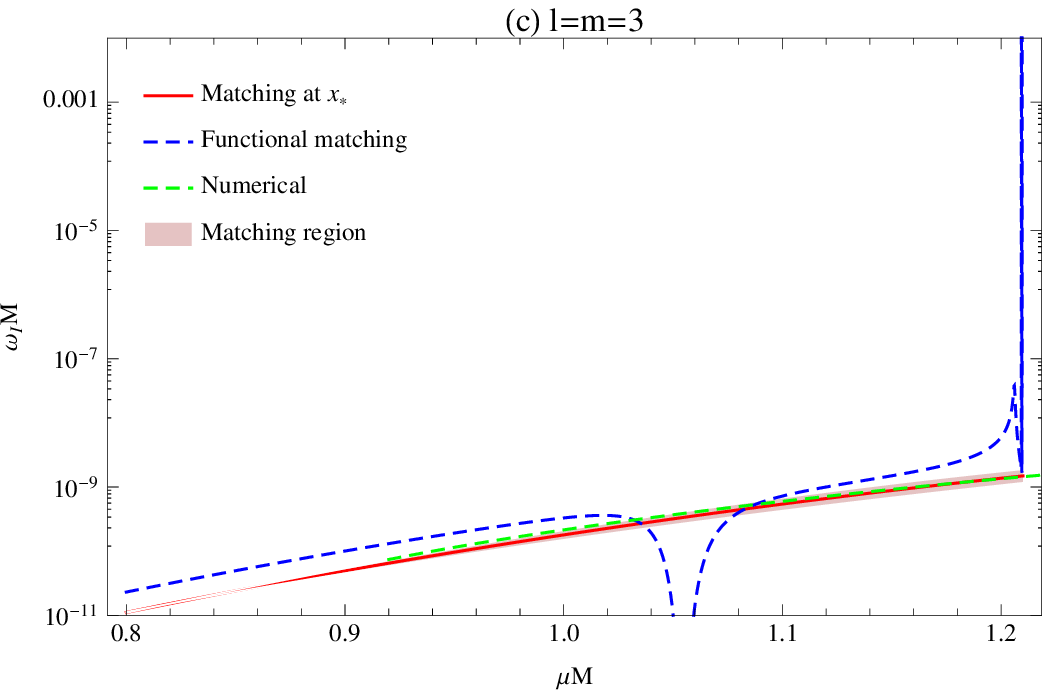, width=9.31cm}
	\caption{Results obtained for the imaginary part of the mode frequency using the point matching method for $l=m=1,~2$ and $3$, with $n=0$, along with the uncertainty associated with the choice of match point. Also shown are the results obtained using the functional matching method and the numerical solutions of \cite{Dolan:2007mj} for $a=0.999 M$. }
		\label{Figure 4}}

As one may observe in the three plots in Figure 4, the match point procedure gives a smooth evolution of $\omega_I M$ as a function of $\mu M$ and does not exhibit any peak-like features as those obtained using the functional matching. For the first three multipoles, these results are in good agreement with the numerically obtained values for $a=0.999M$, although they differ significantly close to the endpoint of the superradiant spectrum. In fact, the endpoint corresponds to $\varpi=0$, which is extremely sensitive to the black hole's spin, extending the superradiant spectrum to larger $\mu$ as one approaches extremality. 

For $l=m=1$, both the numerical and functional matching results (away from the pole) lie within the shaded region and are in good agreement with the values obtained from point matching at $x_*$, showing that in this case the latter is extremely close to the optimal match point. The upper (lower) limit of the shaded region corresponds to matching at $x_{min}$ ($x_{max}$), mainly due to the fact that the Hydrogen-like function $\psi_{II}(x)$ has a larger overlap with the ergoregion if the matching is implemented closer to the horizon. The maximum growth rate for matching at $x_*$ occurs for $\mu M\simeq 0.454$, for which $\omega_I M\simeq 1.49\times 10^{-7}$, in good agreement with the numerical results of \cite{Cardoso:2005vk, Dolan:2007mj}, despite the difference in the black hole's angular momentum and the uncertainty in the choice of the match point. These results clearly show that the four orders of magnitude discrepancy between the results of \cite{Hod:2009cp} and \cite{Cardoso:2005vk, Dolan:2007mj} are unphysical. 

For $l=m=2$ and $3$, Figure 4 shows only a small range of $\mu M$, mainly because for large masses the number of poles in the functional matching condition becomes quite large, which precludes any decent comparison with the point matching technique. For the mass values shown in Figures 4(b) and 4(c), one observes a significantly larger discrepancy between the two analytical techniques away from the poles with respect to the $l=m=1$ case, with the point matching results lying much closer to the corresponding numerical curves. This suggests that the approximations involved in deriving the functional matching condition fail for these multipoles even for $\mu M$ away from the expected poles, so that this method does not give reliable results in this case.

To get a better comparison between the point matching and the numerical methods, we plot in Figure 5 the results for the first three multipoles with $l=m$ and $n=0$, including values for matching at $x_*$, the associated uncertainty for the whole matching region and also the numerical curves for $a=0.99M$ and $a=0.999M$ obtained in \cite{Dolan:2007mj}. 

\FIGURE{
		\epsfig{file=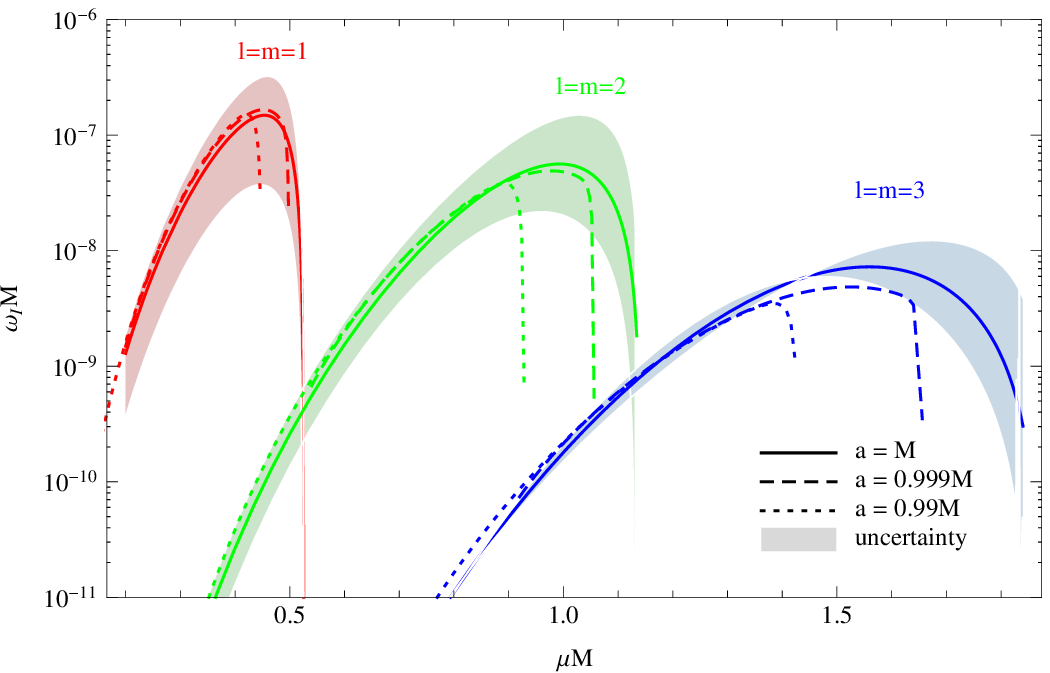}
		\caption{Results for the imaginary part of the mode frequency obtained using the point matching technique for the first three multipoles with $l=m$ and $n=0$, including the corresponding numerical solutions of \cite{Dolan:2007mj} for $a=0.99M$ and $a=0.999M$.}
		\label{Figure 5}}

As expected, the maximum growth rate decreases for larger values of $l=m$, as in these cases the overlap of the radial function with the ergoregion becomes smaller. The agreement between the point-matching and the numerical curves is extremely good, apart from the vicinity of the endpoint where one observes a large discrepancy. In fact, these deviations are larger for the higher multipoles, which is expected as the sensitivity of the condition $\varpi=0$ to the value of $a$ is much greater for larger $m$. This can be readily seen by comparing the two numerical curves for $a=0.99M$ and $a=0.999M$ for each multipole.

The spread in the values of $\omega_IM$ due to the size of the matching region is larger in the neighborhood of the maximum growth rate for the first two multipoles. As discussed earlier, this can be seen as an effective uncertainty associated with the point matching technique, although one should take into account that both $V_I(x)$ and $V_{II}(x)$ are not good approximations to the radial potential close to the limits of the matching region. Also, a larger spread in the values of $\omega_IM$ also indicates a wider overlap region, ensuring that both $\psi_I(x_*)$ and $\psi_{II}(x_*)$ are good approximations to the exact radial function at $x_*$, which is confirmed by the good agreement with the results for $a=0.999M$ up to the value of $\mu M$ giving the maximum growth rate.

On the other hand, the results for $l=m=3$ exhibit a more significant deviation from the numerical ones for small $\mu M$. This is somewhat puzzling, as the uncertainty due to the choice of the match point is negligible in this region. To better investigate these results, we have used the spectrum computed at the match point $x_*$ to determine the boundaries of regions I and II for this multipole. The separation of the two regions for this particular case is illustrated in Figure 6.

\FIGURE{
		\epsfig{file=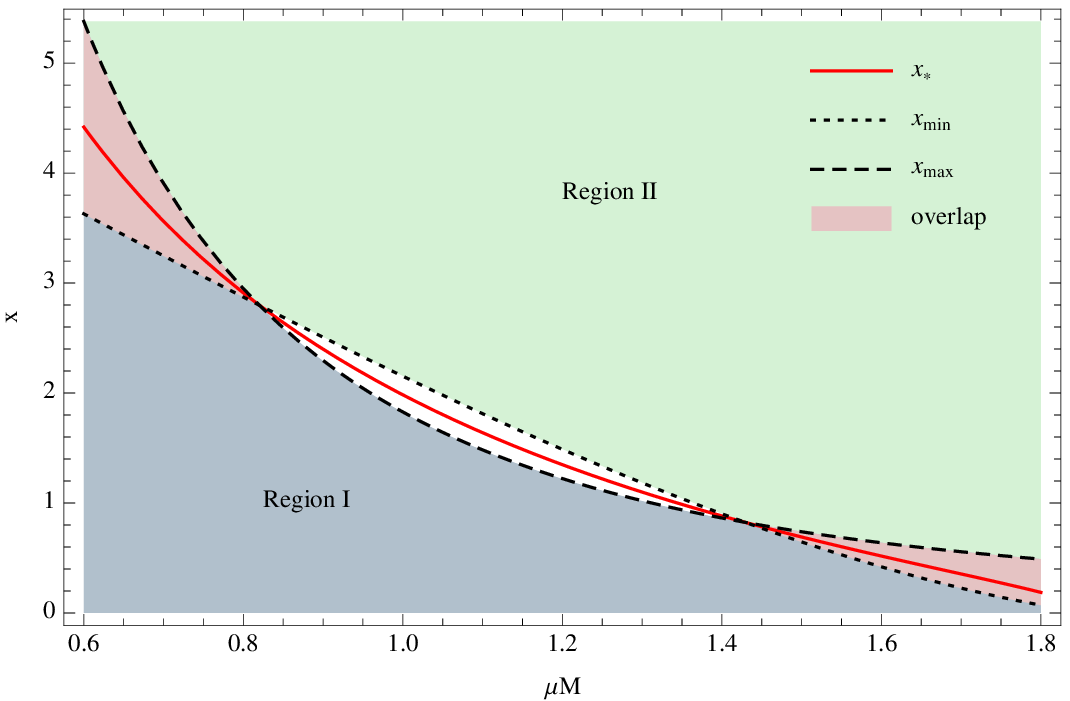}
		\caption{Matching region for $l=m=3$ and $n=0$ as a function of the scalar mass. The dotted and dashed curves correspond to $x_{min}=-4\varpi$ and $x_{max}=(2q)^{-1}$, the lower and upper limits of regions II and I, respectively. The solid line corresponds to the trial match point  $x_*=\sqrt{-2\varpi/q}$.}
		\label{Figure 6}}

As one can easily conclude from this figure, there is no overlap region for  $0.8\lesssim\mu M\lesssim1.4$, which explains the more significant discrepancy for $l=m=3$ with respect to the numerical results in this part of parameter space. For such values of $\mu M$, $x_{min}$ is only slightly larger than $x_{max}$, so that the spread in the corresponding values of $\omega_I M$ is quite small. One has to take into account, however, that some part of this deviation is also due to the non-extremality of the numerical results, a fact that precludes a more accurate determination of the effects of the absence of a matching region. Nevertheless, we may safely conclude that, despite this difference, the point matching at $x_*$ gives a good approximation to the numerically obtained values of $\omega_I M$.


\section{Conclusions and outlook}

In this work, we have analyzed the spectrum of superradiant bound states of a massive scalar field in the background of an extremal Kerr black hole. We have obtained a defining equation for this spectrum using two different techniques, both based on the assumption that one can separate the exterior of the black hole into two overlapping regions where the radial equation is exactly solvable. We have then computed the numerical roots of these conditions in order to determine the growth rate of the superradiant instability as a function of mass of the scalar field.

The first technique, analogous to the one used in earlier works such as \cite{Furuhashi:2004jk} and \cite{Hod:2009cp}, is based on matching the functional forms of the two solutions in their common domain of validity. We have shown that, although this technique is valid for a large range of scalar masses, it exhibits anomalous points which correspond to poles in the matching condition and in the vicinity of which the approximations involved do not hold. For $l=m=1$, one finds two peak-like structures in the vicinity of one of such points, which had been interpreted in {\cite{Hod:2009cp} as a physical enhancement of the growth rate of the instability, disagreeing with earlier numerical computations \cite{Cardoso:2005vk, Dolan:2007mj} by four orders of magnitude. 

As an alternative to the functional matching method, we propose matching the two functions and their derivatives at a single point within the overlap region. Choosing the match point at the geometric mean of the limits of this region, $x_*=\sqrt{-2\varpi/q}$, we obtain a good agreement with the numerical results of \cite{Dolan:2007mj}. The non-extremality of the latter precludes a more accurate comparison of the results obtained using both techniques close to the endpoint of the superradiant spectrum at $\omega=m\Omega$, but the agreement for small $\mu M$ suggests that the results obtained for $x_*$ can be trusted for the whole range of scalar masses. In fact, one could use the numerical results for small masses  to better calibrate the choice of the match point, although the results obtained for $x_*$ are sufficient to illustrate the main properties of the scalar superradiant spectrum.

This method also reveals no anomalous points but only a smooth increase of the growth rate with the scalar mass, with the expected sharp decrease close to the endpoint. Despite the absence of a precise prescription for choosing the match point, this method has a significant advantage over the functional matching method, as it allows a semi-analytical study of the instability over a large range of masses.

We have analyzed the first three multipoles with $l=m$, observing the expected decrease in the maximum growth rate of the instability with increasing $l$. We have obtained a maximum value $\omega_I M\simeq 1.49\times 10^{-7}$  for $\mu M\simeq 0.454$ in the $l=m=1$ case, in close agreement with the results of \cite{Cardoso:2005vk, Dolan:2007mj}.

For pions, this corresponds to a time scale for the development of the instability of $\tau=1.43\times10^{-17}$ sec in a Kerr black hole with $M=8.65\times10^{11}$ kg. This is only about $17\%$ of the pion's mean lifetime, which suggests that, even for small primordial black holes, pions cannot extract sufficient energy and angular momentum from the black hole via the superradiant instability. The proposed (string) axion-black hole bomb \cite{Arvanitaki:2009fg} remains, however, a viable possibility for astrophysical black holes and it is worth investigating the rich phenomenology associated with it.

Both the functional and point matching procedures depend crucially on the existence of an overlap between the near and far regions, which for this analysis corresponds to the condition $-8\varpi q\ll1$. This implies that the matching methods are valid either close to the endpoint of the superradiant spectrum, where $\varpi=\omega-\Omega\ll1$, or for slightly bound states, for which $q=\sqrt{\mu^2-\omega^2}\ll1$. While for the lowest multipoles at least one of these conditions is satisfied for the whole range of scalar masses, in the case $l=m=3$  there is a region of parameter space where no overlap between the two regions exists, effectively rendering any matching procedure inconsistent. This is mainly due to the fact that the lower limit of region II, $x_{min}=-4\varpi$, significantly increases for larger azimuthal number $m$. We cannot {\it a priori} quantify this increase, as it depends on the mode frequency, but one should expect this to be a generic feature of higher multipoles. This does not have, however, important phenomenological consequences, as the growth rate of the superradiant instability is significantly smaller for these multipoles. Also, despite the absence of an ovelap region, the difference between the point matching results and the numerical results of \cite{Cardoso:2005vk, Dolan:2007mj} is small for a large range of scalar masses, so that the former may be sufficiently accurate for most phenomenological purposes. 

The existence of an overlap region for all the values of $\mu M$ considered for $l=m=1$ ensures the applicability of the point matching procedure in this case. On the other hand, the analysis of  \cite{Hod:2009cp} is restricted to the regime $\omega\sim m\Omega\sim\mu$, precluding the use of the matching procedure for small scalar masses. This difference arises from the distinct definitions of the regions I and II (or near and far regions, respectively) used in this work and in \cite{Hod:2009cp}, with the extremality condition playing an important role in this case.

It is worth mentioning that the results obtained in this work, as well as the numerical analysis in \cite{Cardoso:2005vk, Dolan:2007mj}, seem to be in conflict with those of \cite{Strafuss:2004qc}, as pointed out in \cite{Hod:2009cw}. In this numerical analysis of the scalar superradiant instability, the authors estimated a growth rate $\omega_I M=2\times 10^{-5}$ for $\mu M=0.25$, which is two orders of magnitude larger than the maximum value obtained in the present work. The numerical method used in \cite{Strafuss:2004qc} is, however, fundamentally different from the continued-fraction method employed in \cite{Cardoso:2005vk, Dolan:2007mj}. In particular, it computes the time evolution of a particular scalar mode in the Kerr background rather than determining the spectrum of superradiant bound states. Moreover, the selected mode has a frequency $\omega=\mu$ which, as discussed in the present and in earlier analyses, does not correspond to a scalar bound state. Although our results show that the frequency of bound states is very close to the scalar mass $\mu$ when the latter is small, as explicitly obtained in Eq. (\ref{H_spectrum}), they also suggest that small deviations in the real part of the spectrum may produce large (unphysical) effects in the growth rate of the instability. Also, the above mentioned value is derived under the assumption that the mode is growing exponentially, while the estimated e-folding time is in fact much larger than the numerically sampled interval. Thus, the claim in \cite{Hod:2009cw} for the existence of a regime where the superradiant growth rate is much larger than the results obtained in this work is, in our opinion, unjustified and involves a non-trivial extrapolation of the results obtained in \cite{Strafuss:2004qc}.

In the comment to the analysis described in this work of \cite{Hod:2009cw}, the authors of {\cite{Hod:2009cp} explicitly show that all gamma functions involved in their (functional) matching condition are finite for values of $\mu M$ in the region corresponding to one of the peaks obtained for $l=m=1$. This is in agreement with the results obtained in this work, where $\Gamma[2-2\beta]$ exhibits a pole in between these peaks but is finite in the peak regions, as shown in Figure 3. However, as discussed in Section 3, the existence of a pole in the vicinity of the peaks precludes the use of the form in Eq. (\ref{U_limit1}) for the $|z|\rightarrow0$ limit of the hypergeometric function $U(a,b,z)$ required for functional matching. One should use instead the form in Eq. (\ref{U_limit2}), which introduces logarithmic corrections to the power-law form of the solutions in regions I and II. These corrections then invalidate the matching condition in Eq. (\ref{matching1}) and in fact preclude the use of the functional matching procedure in the vicinity of the pole. Although one cannot {\it a priori} determine the values of $\mu M$ for which these logarithmic corrections are significant, the absence of peak- and pole-like features in both the numerical and point matching results, which do not depend on approximate expressions for $U(a,b,z)$, shows that the functional matching procedure does not give the correct values for the superradiant growth rate in a significant range of scalar masses. This is particularly evident for higher multipoles, as illustrated in Figure 4.

Although the near and far regions are constructed in different ways in the present analysis and in that of  {\cite{Hod:2009cp}, these problems are inherent to the functional matching technique and must be taken into account in all applications of this procedure. Furthermore, the observed peak-like structures do not coincide exactly with those obtained in {\cite{Hod:2009cp}, occurring in the present analysis for smaller values of the scalar mass. Given that distinct analyses exhibit similar features at different points in parameter space, this further suggests that both peaks are simply the product of unsuitable approximations and should not be interpreted as physical enhancements of the superradiant growth rate.

An extension of the point matching technique for $a<M$, although outside the scope of this work, would be useful for a better comparison with the numerical growth rates obtained in \cite{Dolan:2007mj} close to the endpoint of the superradiant spectrum, as the continued-fraction method used in the latter work is designed for non-extremal black holes. Also, realistic black holes can only be spun up up to $a\simeq0.998M$, as estimated by Thorne in \cite{Thorne:1974ve} taking into account the effects of radiation from the surrounding accretion disks. The extremal black hole bomb is nevertheless a very good approximation to the latter case, only differing significantly close to the endpoint of the spectrum, where the growth rate becomes very small.

In spite of its limitations, the simple point matching method eliminates most of the shortcomings of other analytical computations of the superradiant spectrum and gives the best agreement with the numerical continued-fraction method. Despite its semi-analytical nature, as it requires solving the non-polynomial matching condition Eq. (\ref{match_point}), it is considerably simpler than a full numerical solution and gives a good theoretical insight on the form of the bound state radial functions. This is crucial for several phenomenological purposes, in particular for determining the backreaction of the axionic clouds formed via superradiant emission around astrophysical black holes. This may, for example, affect the gravity wave signal produced by inspiralling companions, which may lie within the reach of future observatories such as LISA\footnote{I thank John March-Russell for calling my attention to this fact.} \cite{Arvanitaki:2009fg}.

The superradiant instability is extremely important in understanding the stability of rapidly-rotating black holes, at the same time providing a unique astrophysical probe of high-energy particle physics beyond the Standard Model, and we hope that this work motivates further exploitations of the rich phenomenological aspects of this mechanism.

\section*{Acknowledgments}

I would like to thank John March-Russell for interesting discussions on this topic and useful suggestions, as well as for comments on this manuscript. I would also like to thank Sam Dolan for discussions and for kindly furnishing the results of his numerical computations. JGR is supported by FCT (Portugal) under the grant SFRH/BD/23036/2005. This work was partially supported by the EU FP6 Marie Curie Research and Training Network ``UniverseNet" (MRTN-CT-2006-035863).



\end{document}